# An Overview of the JPEG AI Learning-Based Image Coding Standard

Semih Esenlik, Yaojun Wu, Zhaobin Zhang, Ye-Kui Wang,
Kai Zhang, *Senior Member, IEEE*, Li Zhang, *Senior Member, IEEE*
João Ascenso, *Senior Member, IEEE,* Shan Liu, *Fellow, IEEE*

*Abstract*—JPEG AI is an emerging learning-based image coding standard developed by Joint Photographic Experts Group (JPEG). The scope of the JPEG AI is the creation of a practical learning-based image coding standard offering a single-stream, compact compressed domain representation, targeting both human visualization and machine consumption. Scheduled for completion in early 2025, the first version of JPEG AI focuses on human vision tasks, demonstrating significant BD-rate reductions compared to existing standards, in terms of MS-SSIM, FSIM, VIF, VMAF, PSNR-HVS, IW-SSIM and NLPD quality metrics. Designed to ensure broad interoperability, JPEG AI incorporates various design features to support deployment across diverse devices and applications. This paper provides an overview of the technical features and characteristics of the JPEG AI standard.

## I. INTRODUCTION

IMAGE coding standards play an increasingly important role in diversified multimedia applications and services ranging from cloud storage, visual surveillance, media distribution to 360º photo sharing and virtual reality applications. The most widely deployed image coding standard to date is JPEG 1 (ISO/IEC 10918), introduced in 1992, with its latest version released in 1994 [1]. Since then, the Joint Photographic Experts Group (JPEG) has developed many other successful image coding standards including JPEG 2000 [2], JPEG XS [3] and JPEG XL [4]. In the meantime, the Motion Picture Experts Group (MPEG) has created video coding standards such as Advanced Video Coding (AVC) [5] High Efficiency Video Coding (HEVC) [6] and Versatile Video Coding (VVC) [7] that include profiles targeting image coding applications.

In the last decade, machine learning algorithms, such as deep neural networks (DNNs), have become the state-of-the-art for many high-level computer vision tasks like image classification, semantic segmentation, and face recognition, as well as for low-level image processing tasks like image denoising, super-resolution and enhancement [8]. Likewise, significant progress has also been made in the field of learning-based image compression [9]-[14]. Recent research [11][14] demonstrate that learning-based image compression can outperform the VVC standard subjectively and objectively in terms of both MS-SSIM [15] and peak signal-to-noise ratio (PSNR) metrics, making the learning-based image compression technology a strong candidate for the next generation of image coding standard.

The key factors driving progress in machine learning are the advancement of processing power and the increasing availability of hardware accelerators, such as graphics processing units (GPUs) and neural processing units (NPUs). Nowadays GPUs or NPUs are commonly found on servers, PCs and even mobile devices, and are used as AI accelerators for both training and inference. The growing presence of AI accelerators on consumer devices has made the implementation of efficient neural network-based multimedia applications more straightforward.

TABLE I - JPEG AI PARTS

| Part | Name | ITU-T Recommendation \| ISO/IEC International Standard |
|---|---|---|
| Part 1 | Core coding systems | T.JPEG-AI \| 6048-1 |
| Part 2 | Profiling | T.JPEG-AI-P \| 6048-2 |
| Part 3 | Reference software | T.JPEG-AI-RS \| 6048-3 |
| Part 4 | Conformance | T.JPEG-AI-C \| 6048-4 |
| Part 5 | File format | T.JPEG-AI-FF\| 6048-5 |

Recognizing the maturity of learning-based image coding technology and the market's readiness, JPEG launched the JPEG AI project to develop a new image coding standard with learning-based coding technology. In [8] and [16], the strategic vision, foundational motivations, and initial standardization efforts behind JPEG AI were described, setting the stage for a detailed overview of the finalized JPEG AI standard presented here.

The JPEG AI standard adopts several technical features to enable efficient, widespread implementation even on devices with limited computational power and builds on the variational autoencoder with hyperprior architecture [10]. The document on JPEG AI Use Cases and Requirements [17] specifies both essential and desirable requirements that guided the development of the JPEG AI standard. All essential

Semih Esenlik, Zhaobin Zhang, Ye-Kui Wang, Kai Zhang, Li Zhang are with Bytedance Inc., 8910 University Center Ln, San Diego, CA 92122, USA (e-mails: {semih.esenlik, zhaobin.zhang, yekui.wang, zhangkai.video, lizhang.idm}@bytedance.com)

Yaojun Wu is with Bytedance China, Beijing, 100080, China (e-mail: wuyaojun@bytedance.com)

João Ascenso is with Instituto de Telecomunicações - Instituto Superior Técnico, University of Lisbon, 1049-001, Lisbon, Portugal (e-mail: joao.ascenso@lx.it.pt)

Shan Liu is with Tencent Media Lab, 2747 Park Blvd, Palo Alto, CA 94306 USA (e-mail: shanl@global.tencent.com)

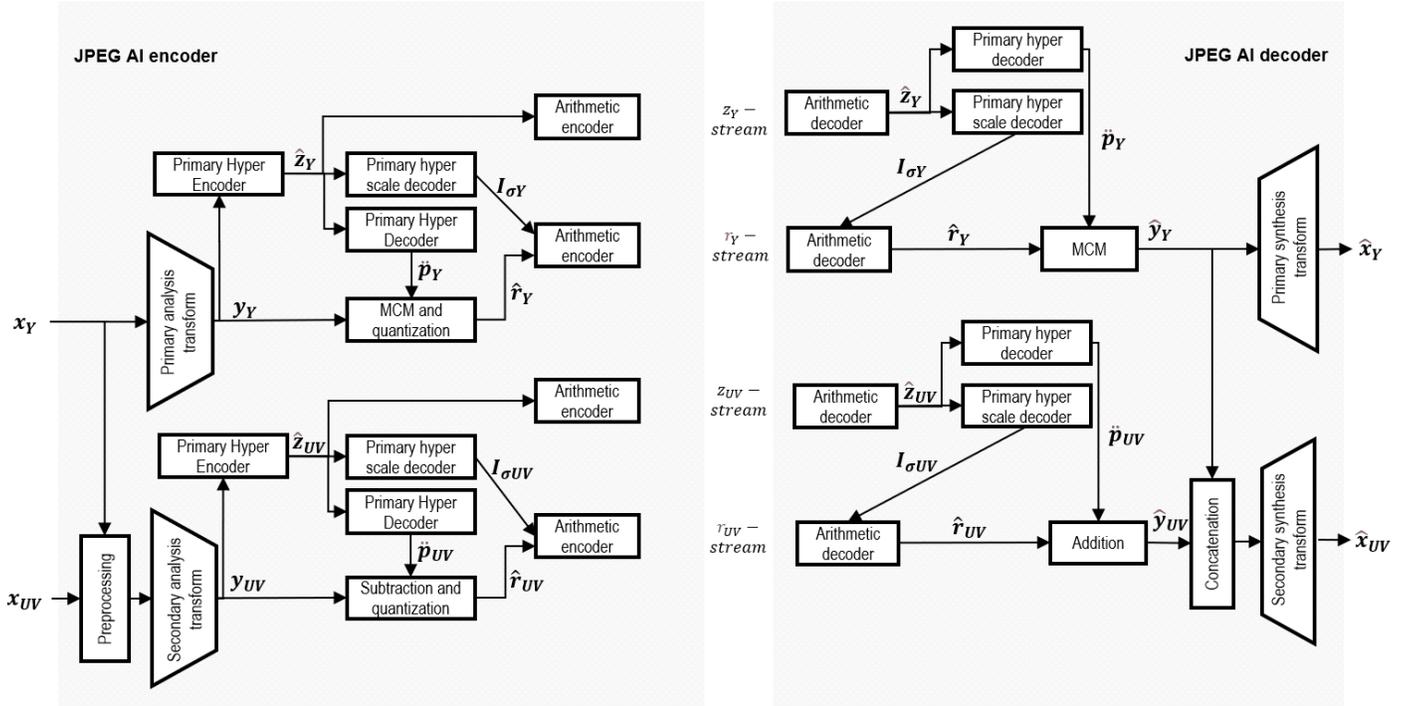

*Figure 1. Framework and the neural network-based components of the JPEG AI encoder and decoder.*

requirements have been fully addressed in the final version, ensuring high-fidelity image reconstruction and high compression efficiency. Throughout the development process, significant emphasis was placed on the reproducibility of the reconstructed images and the ease of implementation across diverse hardware and software environments. This approach enables interoperability and facilitates deployment on a wide range of devices. Other mandatory requirements include synthetic content coding, as well as progressive decoding. Desirable requirements have the potential to expand the application scope. The first version of JPEG AI supports several desirable features, including wide-gamut color coding, multiple color representations, thumbnail image coding, a low complexity profile, region of interest coding, as well as coding based on tiles and regions that enable spatial random access. The spatial random-access functionality refers to the capability to transmit, decode, and display only a spatial subset of an entire coded picture.

The first version of JPEG AI, developed jointly by International Electrotechnical Commission (IEC), the International Organization for Standardization (ISO), and the International Telecommunication Union (ITU), consists of five parts, with their designations listed in Table I. Part 1 specifies the normative algorithms required for parsing the codestream and reconstructing the decoded image. Part 2 defines profiles and levels, i.e. sets of constraints on the codestream and reconstruction process for efficient implementation across various applications. Part 3 specifies the reference software; Part 4 specifies requirements for conformance and Part 5 specifies the file format.

JPEG AI is built upon the multi-branch decoding framework, wherein a single codestream can be decoded to obtain multiple reconstructions, with different complexity and quality tradeoffs, to facilitate the capability of decoding codestreams in a wide range of devices, and the applicability in a wide range of applications. More specifically, after decoding of the codestream via entropy decoding to obtain the quantized residual samples and reconstruction of the latent samples, three synthesis (or inverse) transforms are specified in Part 1 of JPEG AI, each one can generate a reconstructed image. Furthermore, Part 4 of JPEG AI, which is under development at the time of writing, considers the possibility that a decoder is conformant to the standard without having bit-exact reconstruction. The possibility of decoding the same codestream using up to three different synthesis transforms (subject to restrictions that can be imposed by encoder), together with the possibility of standard compliance with a certain leeway, gives vendors the maximum freedom to optimize implementations in a manner appropriate to specific device capabilities and applications.

Like the previous ITU-T and ISO/IEC video coding standards, JPEG AI standardizes the codestream structure, the syntax and the processes required to generate the decoded picture. To assist the industry community in learning how to use the standard, the standardization effort includes not only the development of a text specification document, but also a reference software as an example of how JPEG AI image can be encoded and decoded. The reference software has been used as a research tool for the internal work of the committee during the design of the standard. It can also be used as a general research tool as well as a reference point for product implementations.

This paper is organized as follows. Section II highlights the history and the status of JPEG AI standardization. Section III and IV summarize the design philosophy, coding design and structure of JPEG AI. Section V explains the codestream organization and the high-level syntax design. The JPEG AI coding technology is then described in greater detail in Section VI. Objective and subjective experimental results are provided in Sections VII to demonstrate the performance of JPEG AI codec. Sections VIII, IX, X and XI discuss the profile and level design, conformance specification, reference software and file format of the JPEG AI standard. Finally, Sections XII and XIII provide conclusions and future directions. The Appendix section provides details of the JPEG AI subnetworks, together with definitions of functions and operators used in the manuscript. This overview provides a high-level summary of the JPEG AI standard, and readers are encouraged to refer to the technical specifications in [18][19][20][21][22] for more details.

## II. HISTORY AND STATUS OF JPEG AI STANDARDIZATION

The JPEG AI project was established in 2019 by JPEG, i.e., the Joint Photographic Experts Group (ISO/IEC JTC 1/SC 29/WG 1), to address a set of requirements including standard reconstruction for human visual consumption, computer vision and image processing tasks with a series of standards. The first version of the standard puts the focus on the reconstruction for human visual consumption task, addressing many key requirements such as significant compression efficiency improvement over coding standards in common use, hardware and software implementation-friendly encoding and decoding, and reconstruction reproducibility on different platforms among others [17].

During the early stage of standardization, JPEG AI established the Common Test and Training Conditions (CTTC) [23] and issued a Call for Proposals (CfP) [24] in January 2022. The evaluation metrics for JPEG AI were chosen based on the analysis of objective distortion metrics in the context of learning-based image compression [25]. As a result, seven metrics were used during the development of JPEG AI: MS-SSIM [15], VIF [26], FSIM [27], VMAF [28], NLPD [29], PSNR-HVS-M [30], and IWSSIM [31]. Among the seven responses to the CfP for the standard reconstruction task, two stood out [32][33] and were selected to be combined as the basis for the Verification Model Under Consideration (VMUC) [34]. The harmonization of these two proposals was completed at the JPEG meeting held in October 2022, leading to the establishment of the Verification Model (VM) of JPEG AI [35].

The reference software VM was used to evaluate the effectiveness of proposed coding methods during the development of JPEG AI. Although VM includes an example implementation of all the algorithms described in [18] for demonstration and verification purposes, this reference software is rarely used directly in practical applications. Since the target of JPEG AI is broad deployment capability in a wide range of consumer devices, the main components of JPEG AI have been implemented and tested on various devices during the development [36]-[38] with the purpose of identifying issues related to complexity and interoperability. Reports in [36] and [37] indicated that a $1024 \times 1024$ image could potentially be decoded in under 20ms on a smartphone that is already on the market, and in [38] it was reported that a 4K image could be decoded in approximately 190ms. Although these preliminary implementations are far from the heavily optimized versions found in final products, they helped identify potential issues and validate the practicality of the JPEG AI standard.

## III. JPEG AI DESIGN PHILOSOPHY

The design philosophy of JPEG AI prioritizes widespread adoption, interoperability, and efficient implementation across both current and future mainstream devices. Several key considerations guided its development. A central feature is multi-branch decoding, where the same codestream can be decoded using up to three different synthesis transforms. These provide distinct complexity–efficiency trade-offs (also known as operating points) for implementers while maintaining full interoperability. To further support efficient deployment, JPEG AI defines a bit-exact conformance point only at the end of the entropy decoding pipeline–specifically after the arithmetic decoding stage. Outputs from subsequent processes may diverge slightly from the reference, giving implementers flexibility in how they realize those steps. The concepts of multi-branch decoding and conformance specification are detailed further in the paper.

While JPEG AI is fundamentally based on the Variational Autoencoder (VAE) architecture, several enhancements were introduced to optimize practical implementation. Two of the three operating points are limited to essential neural network layers such as ReLU, ReLU6, and convolution, as these are widely supported by parallel processing accelerators on the market. Only the high operating point (HOP) includes more complex layers, as it targets high-end and future hardware. Additionally, the computational complexity, layer configuration, and memory operations (like pixel shuffling) in the simplified and base operating points were carefully chosen to ensure compatibility with existing devices, a fact confirmed through demonstration implementations during standard development.

## IV. JPEG AI CODING DESIGN AND STRUCTURAL OVERVIEW

The framework of the JPEG AI encoder and decoder are illustrated in Fig. 1. An encoding algorithm that produces a JPEG AI-compliant codestream typically follows the steps described below. JPEG AI commonly employs a YCbCr color space with 4:2:0, 4:2:2, or 4:4:4 color space subsampling, which separates color representation into three components: Y, Cb, and Cr. Samples of Y are organized into the three-dimensional primary component tensor $x_Y[1, H, W]$, while samples of Cb and Cr are organized into the secondary component tensor $x_{UV}\left[2, \left\lceil\frac{H}{c_v}\right\rceil, \left\lceil\frac{W}{c_h}\right\rceil\right]$, wherein $c_v$ and $c_h$ are the chroma subsampling factors of the coded picture, and the $H, W$ are the height and width of the image, respectively. The primary component tensor $x_Y$ is input to primary component analysis transform to obtain the latent representation $y_Y$. The secondary component tensor $x_{UV}$ is preprocessed in combination with $x_Y$ before being input to the secondary component analysis transform, resulting in the latent tensor $y_{UV}$. In the encoder, the primary and secondary components are processed independently, with the sole exception being the preprocessing stage at the start of the secondary component analysis transform.

Following the analysis transformation, $y_Y\left[160,\left\lceil\frac{H}{2^6}\right\rceil,\left\lceil\frac{W}{2^6}\right\rceil\right]$ and $y_{UV}\left[96,\left\lceil\frac{H}{2^6}\right\rceil,\left\lceil\frac{W}{2^6}\right\rceil\right]$ are input to the primary and secondary hyper encoder networks, respectively, output of which are quantized with uniform quantization to obtain the quantized hyper tensors $\hat{z}_Y\left[160,\left\lceil\frac{H}{2^6}\right\rceil,\left\lceil\frac{W}{2^6}\right\rceil\right]$, $\hat{z}_{UV}\left[96,\left\lceil\frac{H}{2^6}\right\rceil,\left\lceil\frac{W}{2^6}\right\rceil\right]$. An arithmetic encoding process with predefined probability tables is used to encode the quantized hyper tensors into substreams $z_Y - stream$ and $z_{UV} - stream$.

After that, $\hat{z}_Y$ is input to the primary hyper decoder and hyper scale decoder networks to obtain the primary prediction tensor $\ddot{p}_Y\left[640,\left\lceil\frac{H}{2^5}\right\rceil,\left\lceil\frac{W}{2^5}\right\rceil\right]$ and the variance tensor in the logarithm domain $I_{\sigma Y}\left[160,\left\lceil\frac{H}{2^4}\right\rceil,\left\lceil\frac{W}{2^4}\right\rceil\right]$, respectively. A multistage context model (MCM)-based prediction and quantization scheme is employed to obtain the quantized primary residual and latent tensors, $\hat{r}_Y\left[160,\left\lceil\frac{H}{2^4}\right\rceil,\left\lceil\frac{W}{2^4}\right\rceil\right]$ and $\hat{y}_Y\left[160,\left\lceil\frac{H}{2^4}\right\rceil,\left\lceil\frac{W}{2^4}\right\rceil\right]$, auto-regressively using $y_Y$ and $\ddot{p}_Y$ as inputs. The MCM in JPEG AI operates in four stages, each stage generating one-fourth of $\hat{r}_Y$, which is used as input for the subsequent stages.

The process of obtaining the quantized residual tensor for the secondary component, $\hat{r}_{UV}$, starts by processing $\hat{z}_{UV}$ through the secondary component hyper decoder and hyper scale decoder networks to generate the secondary component prediction tensor $\ddot{p}_{UV}\left[96,\left\lceil\frac{H}{2^4}\right\rceil,\left\lceil\frac{W}{2^4}\right\rceil\right]$ and the variance tensor in logarithm domain $I_{\sigma UV}\left[96,\left\lceil\frac{H}{2^4}\right\rceil,\left\lceil\frac{W}{2^4}\right\rceil\right]$. From the perspective of the trade-off between computational complexity and coding gain, it was concluded that usage of MCM process for the secondary component is not justified. Instead, $\ddot{p}_{UV}$, which is the output of hyper decoding process is subtracted from $y_{UV}$ followed by rounding to obtain $\hat{r}_{UV}$ according to equation (1).

$$\hat{r}_{UV}[c,i,j] = round(y_{UV}[c,i,j] - \ddot{p}_{UV}[c,i,j]) \quad (1)$$

where $0 \leq c,i,j < 96, \left\lceil\frac{H}{2^4}\right\rceil, \left\lceil\frac{W}{2^4}\right\rceil$.

Finally, arithmetic encoding is employed to encode samples of $\hat{r}_Y$ and $\hat{r}_{UV}$ into substreams $r_Y - stream$ and $r_{UV} - stream$, respectively, assuming zero-mean gaussian distributions with sample variances $I_{\sigma Y}$ and $I_{\sigma UV}$. The resulting sub-streams $z_Y - stream$, $z_{UV} - stream$, $r_Y - stream$ and $r_{UV} - stream$ are encapsulated into codestream segments starting with JPEG AI markers: SOZ (Start of Z-stream), SORp (Start of Residual Stream-primary) and SORs (Start of Residual Stream-secondary). A JPEG AI-compliant codestream shall also include a picture header encapsulated with the PIH (PIcture Header) marker, shall start with an SOC (Start Of Codestream) marker and shall end with an EOC (End Of Codestream) marker, which all together form the mandatory codestream segments. Additionally, the codestream may include supplementary information encapsulated by optional markers.

The encoding process outlined above, along with the details provided in JPEG AI Part 1 [18], serve for informational purposes. In JPEG AI, the codestream structure and syntax are standardized. An encoder is considered conforming to the JPEG AI standard if the codestream generated by it adheres to the syntax structure defined in Part 1, as well as the constraints set forth in Parts 1 and 2 of the JPEG AI standard. This approach allows for a broad range of flexibility in encoder implementations, accommodating applications with diverse requirements.

On the other hand, the decoding process, which converts the codestream into a decoded picture, is normative. A decoder implementation must follow the processes specified in the JPEG AI part 1 standard and obtain reconstructions that are conformant. The decoding process starts with the entropy decoding pipeline. First, the $z_Y - stream$ and $z_{UV} - stream$ are decoded to obtain the primary and secondary quantized hyper tensors, $\hat{z}_Y$ and $\hat{z}_{UV}$, respectively. The hyper scale decoding process is applied using $\hat{z}_Y$ and $\hat{z}_{UV}$ as inputs, resulting in the primary and secondary variance tensors in the logarithm domain, $I_{\sigma Y}$ and $I_{\sigma UV}$, respectively. The variance tensors are used to decode the primary and secondary residual substreams $r_Y - stream$ and $r_{UV} - stream$ through entropy decoding to obtain the quantized residual tensors $\hat{r}_Y$ and $\hat{r}_{UV}$, respectively. $\hat{z}_Y$ and $\hat{z}_{UV}$ are also used by the primary and secondary hyper decoding processes to obtain primary and secondary prediction tensors $\ddot{p}_Y$ and $\ddot{p}_{UV}$.

Decoding of $\hat{r}_Y$ and $\hat{r}_{UV}$ is followed by reconstruction of the quantized latent tensors $\hat{y}_Y\left[160,\left\lceil\frac{H}{2^4}\right\rceil,\left\lceil\frac{W}{2^4}\right\rceil\right]$ and $\hat{y}_{UV}\left[96,\left\lceil\frac{H}{2^4}\right\rceil,\left\lceil\frac{W}{2^4}\right\rceil\right]$. The reconstruction of $\hat{y}_Y$ includes processing the quantized primary residual tensor $\hat{r}_Y$ and the primary prediction tensor $\ddot{p}_Y$ with the MCM submodule, whereas $\hat{y}_{UV}$ is obtained by adding the secondary prediction and quantized secondary residual tensors.

$$\hat{y}_{uv}[c,i,j] = \hat{r}_{uv}[c,i,j] + \ddot{p}_{UV}[c,i,j] \quad (2)$$

where $0 \leq c,i,j < 96, \left\lceil\frac{H}{2^4}\right\rceil, \left\lceil\frac{W}{2^4}\right\rceil$.

Finally, the reconstructed primary component $\hat{x}_Y[0,H,W]$ is obtained by processing $\hat{y}_Y$ with the primary synthesis transform, whereas $\hat{y}_{uv}$ is first concatenated with $\hat{y}_Y$ to obtain $\hat{y}^c{}_{UV}$ according to (3), which is then fed to the secondary synthesis transform to obtain the reconstructed secondary component $\hat{x}_{UV}$.

$$\hat{y}^c{}_{UV}[c,i,j] = \begin{cases} \hat{y}_{UV}[c,i,j], & 0 \leq c < 96 \\ \hat{y}_Y[c,i,j], & 96 \leq c < 256 \end{cases} \quad (3)$$

where $0 \leq i, j < \left\lceil \frac{H}{2^4} \right\rceil, \left\lceil \frac{W}{2^4} \right\rceil$.

In Fig. 1, the neural network-based components of the JPEG AI standard are demonstrated. Additionally, JPEG AI also employs switchable coding tools to adjust the variance tensors ($I_{\sigma Y}$ and $I_{\sigma UV}$), quantized residual tensors ($\hat{r}_Y$ and $\hat{r}_{UV}$), reconstructed latent tensors ($\hat{y}_Y$ and $\hat{y}_{UV}$), and reconstructed image components ($\hat{x}_Y$ and $\hat{x}_{UV}$), which are detailed in Section VI.

Two important design decisions are worth emphasizing in the entropy decoding pipeline. Although the JPEG AI decoding process consists mainly of highly parallelizable components suitable for implementation in the state-of-the-art NPU and GPU architectures, the arithmetic decoding process is a sequential process that might necessitate implementations on general purpose processing units such as CPU. The primary and secondary hyper scale decoding processes, which are interleaved in between two arithmetic decoding processes (i.e. decoding of quantized hyper tensor and residual tensor), are designed to be computationally lightweight to allow efficient implementations of the whole entropy decoding pipeline on a CPU. Secondly, the entropy decoding pipeline employs integer arithmetic neural network operations that can be implemented with 8-bit multipliers and 32-bit registers, which allows bit-exact replication of tensors $\hat{r}_Y$ and $\hat{r}_{UV}$ (and hence $\hat{z}_Y$ and $\hat{z}_{UV}$) on different devices and platforms, contributing to the interoperability of JPEG AI. A theoretical proof is provided in [39] to demonstrate that the overflow is not possible when 32-bit registers used in the whole entropy decoding pipeline.

## V. Codestream organization and high-level syntax

A JPEG AI compliant codestream starts with an SOC marker and ends with an EOC marker. The codestream is divided into codestream segments, each starting with a 2-byte marker, followed by the variable-length coded substream size and the payload. Byte alignment is invoked at the end of both the substream size indication and at the end of the payload. Table II lists the markers, their corresponding hexadecimal code assignments and the encapsulated payload types in JPEG AI standard.

The codestream segments starting with PIH, SOZ, SORp and SORs are mandatory, whereby a JPEG AI compliant codestream is required to include at least one codestream segment starting with each of these markers. More than one segment starting with SORp and SORs might be present in the codestream, provided that the corresponding region index (*region_idx[0]* syntax element in SORp and *region_idx[1]* for SORs) value is different for each segment. The rest of the markers are optional and not required to be present in the codestream.

TABLE II – JPEG AI MARKERS

| Symbol | Code assignment | Payload | Mandatory/ Optional |
|---|---|---|---|
| SOC | 0xff80 | None | Mandatory |
| EOC | 0xff81 | None | Mandatory |
| PIH | 0xff82 | Picture header | Mandatory |
| SOZ | 0xff88 | $z_Y - stream$ and $z_{UV} - stream$ | Mandatory |
| SORp | 0xff89 | $r_Y - stream$ | Mandatory |
| SORs | 0xff8a | $r_{UV} - stream$ | Mandatory |
| TOH | 0xff83 | Tools header | Optional |
| SOQ | 0xff8b | Quality map information | Optional |
| UDI | 0xff8c | User defined information | Optional |
| RDI | 0xff84 | Rendering information | Optional |

### A. Picture header

The picture header is encapsulated with a codestream segment starting with PIH marker and shall be present in JPEG AI compliant codestreams. It includes information pertaining to profiling, picture size, output bit-depth, internal and output subsampling modes, model indices, as well as parameters controlling the operation of the following coding tools: color space conversion, picture partitioning into regions and tiles, multithreaded entropy decoding, Residual and Variance Scaling (RVS), rate adaptation and skip mode.

### B. Tools header

The tools header is an optional codestream segment that includes the control parameters of the switchable coding tools that are applied after the entropy decoding pipeline at the decoder, i.e. after the quantized residual tensors $\hat{r}_Y$ and $\hat{r}_{UV}$ are obtained. Within the tools header, two flags indicate whether Latent Scaling Before Synthesis (LSBS) is enabled for the primary and secondary components, respectively. Following this, the filters header is included that comprise parameters of the post processing filters. Specifically, four enable flags are used to control the usage of the enhancement filtering extensions (EFE) tools, namely the EFE linear filter, Inter Channel Correlation Information (ICCI) filter, EFE nonlinear filter, and Luma Edge Filter (LEF). When any of these filters are enabled, the corresponding control information is included in the tools header. If the tools header codestream segment is absent, all the aforementioned enable-flags are inferred to be equal to 0, indicating that the tools are disabled.

### C. Quality map information

JPEG AI allows signaling different quantization step sizes to code different samples of the quantized residual tensors $\hat{r}_Y$ and $\hat{r}_{UV}$. This functionality, named 3D gain unit in the context of JPEG AI Part 1, utilizes a two-dimensional quality map tensor, which is encapsulated in the quality map information codestream segment. The 3D gain unit provides spatially variable control over the quantization process and is useful in applications like Region of Interest (RoI) coding. The 3D gain unit process is detailed in Subsection VI.I.

### D. User defined information

The User Defined Information (UDI) is an optional codestream segment, encapsulated with the UDI marker. It enables applications of the JPEG AI standard (being an application standard or a proprietary application specification that uses JPEG AI) to include their own information in any format in a JPEG AI compliant codestream.

*E. Rendering information*

The rendering information codestream segment is marked with the RDI (RenDering Information) marker and is optional in JPEG AI compliant codestreams. It carries rendering related information, including 1) properties of the reconstructed picture, such as the color primaries, transfer characteristics, matrix coefficients, image full range flag, and chroma sample location; 2) the color volume (the color primaries, white point, and luminance range) of a display considered to be the mastering display for the content – e.g., the color volume of a display that was used for viewing while authoring the content; 3) upper bounds for the nominal target brightness light level of the picture; and 4) information on dynamic metadata for a picture with high dynamic range.

## VI. JPEG AI IMAGE CODING TECHNIQUES

*A. Separated processing of color components*

As in other image coding standards, the JPEG AI codec doesn't use the RGB color space for encoding. For representing color image signals, JPEG AI typically uses a YCbCr color space with 4:2:0, 4:2:2 or 4:4:4 sub-sampling. This separates a color representation into three components called Y, Cb, and Cr. The Y component, a.k.a. luma, represents illumination, whereas Cb and Cr, known as chroma, represent blue-difference and red-difference. The main reason is that human eye is much more sensitive to luminance (brightness) than to chrominance, but also because YUV reduces correlation between components compared to RGB, rendering the compression to be more efficient. It also allows prioritizing luminance during the training, achieving better RD performance, and is a color space used in many legacy devices.

The chroma subsampling mode is controlled by four flags that are signaled in the picture header, *s_ver_minus1* and *s_hor_minus1* controlling the subsampling format of the intended output picture, and *c_ver_minus1* and *c_hor_minus1* controlling the coded picture used in interal processing, i.e. the layer configuration of the synthesis transform. If internal coding format (i.e. the output of the synthesis transform) is different from the intended output format, a normative upsampling process is applied to generate the reconstructed picture.

As shown in Fig. 1, the JPEG AI codec is divided into primary and secondary component branches, which process the luma (primary) and chroma (secondary) components separately. The luma component is processed independently and the chroma components are processed using information from the luma component at two stages, i.e., the beginning of the analysis transform and the beginning of the synthesis transform. No other information exchange occurs during processing of the chroma components, except during the non-normative post-processing filtering stages.

The approach of separately processing the luma and chroma components may not be ideal in terms of the coding gain, as there are usually significant correlations between the samples of the luma and chroma components. However, the primary motivation for this approach is to reduce computational complexity. During the standardization process, it was found that limiting the cross-component information exchange to just two stages—at the beginning of analysis and the synthesis transform—provides a reasonable trade-off between coding gain and complexity.

Additionally, the feature of decoding the luma component independently can be helpful in monochrome applications, such as computer vision applications that typically use only the luma component. In such cases, decoding of the chroma component can be skipped, further reducing the implementation cost.

*B. Color space conversion*

The model parameters, e.g. weights and biases of the neural network components, which are an integral part of the JPEG AI standard, are trained by first converting the training images into YCbCr color space using the conversion specified in ITU-T Recommendation R.709-6. Though JPEG AI does not mandate a specific color space to represent the primary and secondary components, due to the training assumption, the compression performance might not be optimal if a different internal color space is used for coding. To address a potential color space conversion at the beginning of encoding, JPEG AI employs a mandatory color space conversion at the end of the decoding process, to allow reverting the internal color space to intended output color space. The color space conversion is applied after the whole decoding process, i.e. $\hat{x}_Y$ and $\hat{x}_{UV}$ in equations below are the outputs of the post-processing filters.

Parameters of color space conversion is signaled in the picture header using one of the three signaling modes specified by the syntax element *colour_transform_idx*. When *colour_transform_idx* is equal to 1, no conversion is applied. When it is equal to 0, the reconstructed image is converted using a predefined equation specifying the conversion from YCbCr to RGB color space as in (4).

$$\hat{x}[0,i,j] = \hat{x}_Y[0,i,j] + 1.5748(\hat{x}_{UV}[1,i,j] - 0.5)$$
$$\hat{x}[1,i,j] = \hat{x}_Y[0,i,j] + 1.8556(\hat{x}_{UV}[1,i,j] - 0.5)$$
$$\hat{x}[2,i,j] = \frac{(\hat{x}_Y[0,i,j] - 0.2126 \cdot \hat{x}[0,i,j] - 0.07222 \cdot \hat{x}[1,i,j])}{0.7152}$$
(4)

where $0 \leq i, j < H, W$.

If *colour_transform_idx* is equal to 2, the color space conversion can be defined by the encoder in a flexible way. By signalling a weight martix *a*[3,3] and a bias vector *b*[3] in the picture header, the color space conversion is performed according to (5).

$$\hat{x}[0,i,j] = \hat{x}_Y[0,i,j] \cdot a[0,0] + \hat{x}_{UV}[0,i,j] \cdot a[0,1] + \hat{x}_{UV}[1,i,j] \cdot a[0,2] + b[0]$$
$$\hat{x}[1,i,j] = \hat{x}_Y[0,i,j] \cdot a[1,0] + \hat{x}_{UV}[0,i,j] \cdot a[1,1] + \hat{x}_{UV}[1,i,j] \cdot a[1,2] + b[1]$$
$$\hat{x}[2,i,j] = \hat{x}_Y[0,i,j] \cdot a[2,0] + \hat{x}_{UV}[0,i,j] \cdot a[2,1] + \hat{x}_{UV}[1,i,j] \cdot a[2,2] + b[2]$$
(5)

where $0 \leq i, j < H, W$.

At the end, the output reconstructed picture is scaled and clipped according to the bitdepth parameter signaled in the picture header as follows:

$$clip\left(0, 2^{bitdepth} - 1, \hat{x}[c,i,j] \cdot (2^{bitdepth} - 1)\right) \rightarrow \hat{x}[c,i,j] \quad (6)$$

where $0 \leq c, i, j < 3, H, W$, $bitdepth$ is the output bit-depth signalled in the picture header and $\hat{x}$ is the output of the JPEG AI decoder.

### C. Entropy coding pipeline

The entropy coding pipeline was designed to be rather simple to be able to achieve high throughput and to be implemented in CPU. This can be achieved by using a second-level latent representation (also called hyperprior) that summarizes the primary latent variables used to reconstruct the image. In this case, it is used to predict the distribution (e.g., standard deviation) of the latent codes, making entropy coding more accurate.

The entropy decoding pipeline consists of decoding of hyper tensors $\hat{z}_Y$ and $\hat{z}_{UV}$ from $z_Y - \text{stream}$ and $z_{UV} - \text{stream}$, calculation of variance samples $I_{\sigma Y}$ and $I_{\sigma UV}$, and decoding of quantized residual tensors $\hat{r}_Y$ and $\hat{r}_{UV}$ from $r_Y - \text{stream}$ and $r_{UV} - \text{stream}$ using the variance samples.

In decoding of the hyper tensors, a fixed Cumulative Distribution Function (CDF) is used, represented as a 2D table with dimensions [128, 64], as specified in [18]. Based on this CDF table, a memory-efficient tabulated Asymmetric Numeral System (me-tANS) coder is employed to decode $\hat{z}$ from the codestream. Since the bit cost corresponding to $\hat{z}$ constitutes only a small portion of the overall codestream, both $z_Y - \text{stream}$ and $z_{UV} - \text{stream}$ are encapsulated into a single codestream segment that is marked with SOZ codestream marker. Table II lists the codestream segments used in JPEG AI along with their markers and payload.

The hyper tensors $\hat{z}_Y$ and $\hat{z}_{UV}$ are processed via the hyper scale decoder network to generate $I_{\sigma Y}$ and $I_{\sigma UV}$, which represent the standard deviation tensors to be used in entropy decoding. JPEG AI uses a zero-mean gaussian distribution to model the sample distribution of residual samples. The standard deviation tensors $I_{\sigma Y}$ and $I_{\sigma UV}$ are then used in me-tANS to obtain the residual samples $\hat{r}_Y$ and $\hat{r}_{UV}$. To ensure an efficient entropy decoding process and device interoperability, the hyper-scale decoder is designed with minimal complexity and is quantized to provide identical results across different architectures and devices.

Decoding of the residual samples of tensors $\hat{r}_Y$ and $\hat{r}_{UV}$ from the codestream uses variance tensors $I_{\sigma Y}$ and $I_{\sigma UV}$, and a pre-calculated CDF table that has dimensions of [32, 256]. Each row of the CDF table represents a cumulative distribution. The distribution to be used in decoding of a sample $\hat{r}_Y[c,i,j]$ is selected by using $I_{\sigma Y}[c,i,j]$ as index to the table. A significant increase in entropy decoding throughput is achieved with usage of precalculated distributions (hence the name tabulated ANS). The residual samples for primary component $\hat{r}_Y$ and the secondary component $\hat{r}_{UV}$ are encapsulated in separate codestream segments, that are marked with SORp and SORs markers respectively.

The entropy coder employs a skip mode to bypass the coding of residual samples based on the corresponding standard deviation value. If $I_\sigma[c,i,j]$ corresponding to the residual sample $\hat{r}[c,i,j]$ is smaller than a fixed predefined threshold, the coding of the sample is skipped, increasing the throughput of entropy coding.

With the skip mode, depending on the content, coding of up to 80% of the residual samples might be skipped. Though this brings significant speedup in entropy decoding, during the development of JPEG AI, it is noticed that the fixed threshold used in skip mode may cause visual artefacts in rare cases. To address any such issue, a cube-based skip mode is adopted as a safeguard mechanism, which allows reverting the decision of the skip mode for each 16x16x16 sized partitions of the residual tensor with explicit signaling.

Multithreading can be enabled for codestream segments encapsulating $z_Y - \text{stream}$, $z_{UV} - \text{stream}$, $r_Y - \text{stream}$, $r_{UV} - \text{stream}$ and $q - \text{stream}$ to further increase the throughput in multi-core devices. In this case, the payload is divided into independently decodable substreams, where the substream offsets (i.e. the pointers to starting point of each substream) are signaled at the beginning of the codestream segment. The number of substreams are signaled in the picture header.

At the core of JPEG AI's entropy coding process is a specialized implementation of Asymmetric Numeral Systems (ANS) called me-tANS. Symbols are decoded in a first-in, last-out (FILO) fashion using four precomputed tables as outlined in Algorithm 1. When decoding residual samples $\hat{r}[c,i,j]$, the specific table used is selected based on the value of $I_\sigma[c,i,j]$. For decoding hyper prior samples $\hat{z}[c,i,j]$, the table selection depends on the channel index c.

If the value (and therefore the probability) of a symbol is smaller than a threshold defined by the bound_table, it is decoded directly using the table. If the value equals the threshold, an additional outbound value is decoded from the bitstream, and the final symbol value is calculated as the sum of this outbound value and the threshold. This two-step approach increases throughput for high-probability symbols while preserving the ability to encode symbols with larger values when necessary.

Algorithm 1 describes the core me-tANS procedure for residual sample decoding in a single-threaded context. Stream interleaving and stream concatenation techniques are employed to achieve parallel coding with multiple threads. Even in single-threaded operation, two substreams are interleaved and two separate states are maintained to enhance parallelism—effectively mimicking the behavior of a dual-threaded setup. Thanks to its tabular design, decoding requires only a bitwise OR operation and a simple addition, making the implementation extremely lightweight in terms of computation. Furthermore, as the name suggests, the memory required to store the tables is very limited, typically on the order of 100 KB at the decoder side.

**Algorithm 1** Core me-tANS pseudocode for one thread

**Input:** $I_\sigma[\ ]$, $bitstream$
**Output:** $\hat{r}[\ ]$
**Initialization:**
   Organize $I_\sigma[\ ]$ into 1-D array.
   Move the pointer in the bitstream to the last symbol position. /*the pointer moves backwards*/
    $s \leftarrow$ Parse 8 bits from $bitstream$.
**for** $i = 0$ **to** number of symbols / 4 **do**
   **for** j = 0 **to** 3 **do**
     $\hat{r}[4i+j] \leftarrow$ transition_table_symbol$[I_\sigma[4i+j]][s]$
     $n \leftarrow$ transition_table_nBits$[I_\sigma[4i+j]][s]$
     $value \leftarrow$ parse n bits from $bitstream$
     $s \leftarrow$ (transition_table_stateNext$[I_\sigma 4i+j][s]$|value)
   **end for**
   **for** j = 0 **to** 3 **do**
     **if** $\hat{r}[4i+j]$ + bound_table$[I_\sigma[4i+j]]$ == 0 **do**
       $ind \leftarrow$ Parse 1 bit from $bitstream$
       $m \leftarrow ind$? 2: 15
       $value \leftarrow$ Parse $m$ bits from $bitstream$
       $sign \leftarrow$ Parse 1 bit from $bitstream$
       $\hat{r}[4i+j] \leftarrow$ bound_table$[I_\sigma[4i+j]] + value \times sign$
     **end if**
   **end for**
**end for**
Reorganize $\hat{r}[\ ]$ into 3-D tensor.
**return** $\hat{r}[\ ]$

### D. Latent sample reconstruction

In latent space reconstruction, the quantized latent tensors that feed the synthesis network are obtained. Each latent sample is computed from hyperprior side information and already-decoded neighbors, which allows to reduce the number of bits needed while keeping decoder complexity manageable through multi-stage, partially parallel processing. Thus, spatial dependencies between latent symbols are exploited to improve compression efficiency. This is a key distinction of JPEG AI from many other state-of-the-art learning-based image codecs.

The input of the latent sample reconstruction process are quantized residuals $\hat{r}_Y$ and $\hat{r}_{UV}$ and the hyper tensors $z_Y$ and $\hat{z}_{UV}$. The primary component quantized latent tensor $\hat{y}_Y$ is generated by using the primary component hyper decoder and the Multi-stage Context Modeling (MCM) submodules, whereas the secondary component latent tensor $\hat{y}_{UV}$ is generated solely using the secondary component hyper decoder.

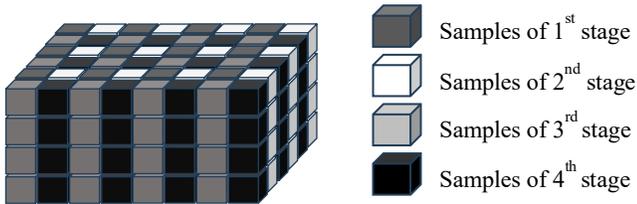

*Figure 2. The checkerboard pattern used in grouping of primary component latent samples.*

Reconstruction primary latent tensor starts with utilizing the hyper tensor $\hat{z}_Y$ to obtain primary prediction tensor $\ddot{p}_Y$. Quantized residual tensor $\hat{r}_Y$ and $\ddot{p}_Y$ are fed to MCM module, where $\hat{y}_Y$ is reconstructed by MCM in 4 stages. In each stage of MCM, one group of latent samples are reconstructed using the corresponding samples of $\hat{r}_Y$, $\ddot{p}_Y$ and all samples of the latent tensor $\hat{y}_Y$ that are already reconstructed. The order of reconstruction of the latent samples is illustrated in Fig. 2. Usage of already reconstructed samples as context improves the coding gain, whereas the number of processing stages imposes further complexity. During the development of JPEG AI, various latent sample reconstruction schemes have been evaluated, with the current design with 4 processing stages achieving the best compromised between complexity and coding efficiency. The structure of each MCM stage is detailed in the Subsection C of the Appendix. The network structures of hyper decoder and hyper scale decoder are detailed in the Subsections A and B of the Appendix.

### E. Decoupled entropy decoding and latent reconstruction

While JPEG AI primarily employs highly parallelizable neural network modules, the arithmetic decoding of a residual substream is still sequential. This might necessitate a different design approach for the entropy decoding pipeline, which would need to be executed predominantly on a CPU — even on devices equipped with a GPU or NPU. One key design element of JPEG AI is decoupling of the entropy decoding pipeline (which outputs $\hat{r}_Y$ and $\hat{r}_{UV}$) from the subsequent latent sample reconstruction process which generates $\hat{y}_Y$ and $\hat{y}_{UV}$. To achieve this, JPEG AI incorporates a hyper decoder module and a hyper scale decoder module for processing quantized hyper tensors: the former is part of latent sample reconstruction, while the latter is part of the entropy decoding pipeline. [40] provides further details of the decoupled entropy decoding and latent reconstruction.

This design offers flexibility for implementing the decoder in real-world products. For instance, the entropy decoding pipeline can be implemented as a dedicated separate engine (e,g, running of CPU), while latent sample reconstruction can be optimized for execution on an NPU or GPU; however, implementation choices are not limited to this configuration.

### F. Synthesis transform

The synthesis transform is very important in JPEG AI since it reconstructs the decoded image. As this stage concentrates most of the codec's computational burden, JPEG AI defines three distinct synthesis transforms, identified by decoderID = 0, 1, 2, to accommodate devices with varying processing capabilities. Each option represents a specific trade-off between reconstruction quality and computational complexity. After the reconstruction of latent tensors $\hat{y}_Y$ and $\hat{y}_{UV}$, one of the three synthesis transformations can be performed to generate a JPEG AI conformant reconstruction. The set of allowable synthesis transforms is further limited in JPEG AI Part 2 [19] depending on the profile. Three synthesis transforms are designed for devices and applications with different capabilities in mind:

1) *decoderID* = 0: Devices with limited parallel processing capability, or applications with tight timing or power constraints. This might primarily target devices such as laptops without neural network acceleration or mid-range and low-end mobile devices.

TABLE III EXPERIMENTAL RESULTS USING JPEG AI COMMON TEST CONDITIONS

|  |  | AVG | MS-SSIM | VIF | FSIM | NLPD | IW-SSIM | VMAF | PSNR-HVS | kMAC/pxl | Decoding Time (ms/megapixel) |
|---|---|---|---|---|---|---|---|---|---|---|---|
| encoderID = 0 | decoderID = 0 | -16.2% | -30.9% | 6.9% | -22.3% | -13.4% | -26.6% | -30.0% | 2.9% | 14 | 285 |
|  | decoderID = 1 | -20.2% | -33.0% | 1.4% | -26.9% | -17.3% | -29.1% | -34.8% | -1.9% | 28 | 266 |
|  | decoderID = 2 | -22.1% | -34.8% | -2.0% | -27.7% | -19.3% | -31.2% | -37.3% | -2.5% | 215 | 323 |
| encoderID = 1 | decoderID = 0 | -14.4% | -30.3% | 9.7% | -20.4% | -11.6% | -25.1% | -29.2% | 6.1% | 14 | 246 |
|  | decoderID = 1 | -19.9% | -33.0% | 1.5% | -26.4% | -16.7% | -28.4% | -35.8% | -0.8% | 28 | 271 |
|  | decoderID = 2 | -27.0% | -37.6% | -8.5% | -34.7% | -23.6% | -33.7% | -42.4% | -8.8% | 215 | 332 |

2) *decoderID* = 1: Devices with moderate parallel processing capability, such as high-end mobile devices with NPU or GPU, or applications with less strict timing or power constraints.
3) *decoderID* = 2: Devices employing high parallel processing capabilities, such as desktop computers with graphic processing units.

Table III presents the BD-rate gain and complexity of the three synthesis transforms comparatively. The network structure of the primary and secondary synthesis transforms with ID 0 and 1 are depicted in figures 6 and 7 in the Appendix section respectively, whereas figure 8 depicts the primary and secondary synthesis transforms for decoderID 2.

The first two operating points (decoderIDs 0 and 1) are significantly different from the third operating point (decoderID=2), as special attention has been given to ensure their implementability on currently available devices. The main distinction between the first and second operating points in their choice of upsampling layers: the first uses 2×2 convolution followed by pixel shuffling, while the second employs 4×4 deconvolution. Although pixel shuffling typically requires less computation than deconvolution, it was found out that implementations of pixel shuffling might be inefficient on certain devices. To address this, JPEG AI defines the second operating point to restrict the use of pixel shuffling to the final layer, offering greater flexibility to implementers. Additionally, the number of channels of the final layers are 32 and 64 in the first and second operating points respectively. This difference makes the first operating point computationally much simpler, making it suitable for deployments where parallel processing acceleration might be limited (e.g. using only CPU).

### G. Residual and Variance Scaling (RVS)

The model parameters that are part of JPEG AI Part 1 are obtained by a training process using a distortion metric that is a weighted combination of MSE and MS-SSIM. On the other hand, seven different metrics are used to evaluate the performance of JPEG AI. RVS is a coding tool that enables adjustment of the tensors $I_{\sigma Y}$, $I_{\sigma UV}$, $\hat{r}_Y$ and $\hat{r}_{uv}$ conditionally on $I_{\sigma Y}$ and $I_{\sigma UV}$ to achieve BD-rate improvements primarily on metrics VMAF, FSIM and NLPD, which are not part of the training loss function. RVS is applied to both the primary and secondary components. In this subsection, only the application of RVS to the primary component is described, as the application to the secondary component is identical.

Two flags, grfs_enable_flag[comp] and rvs_enable_flag[comp] are signalled in the picture header to control application of RVS on the primary and secondary components (comp=0 and comp=1 indicating primary and secondary components, respectively). The RVS process on the primary component starts by computing the average variance tensor $\sigma_Y$ via average pooling of $I_{\sigma Y}$.

$$\sigma_Y[c,i,j] = \left(32 + \sum_{i',j'=0,0}^{7,7} I_{\sigma Y}[c, 8i+i', 8j+j']\right) \gg 6 \quad (7)$$

where $0 \leq c < 160$ and $0 \leq i,j < \left\lceil\frac{H}{128}\right\rceil, \left\lceil\frac{W}{128}\right\rceil$.

Tensor $I_{\sigma Y}$ is padded with value 1411 at the boundaries in the case when the indices exceed the tensor boundaries. 1411 is selected since it represents the average value of the samples of $I_{\sigma Y}$. The $\hat{r}_Y$ and $I_{\sigma Y}$ are then updated according to (8) and (9).

$$I_{\sigma Y}[c,j,k] + T_1\left[modelID, id[c], \sigma_Y\left[c, \left\lfloor\frac{i}{8}\right\rfloor, \left\lfloor\frac{j}{8}\right\rfloor\right]\right] \quad (8)$$
$$\rightarrow I_{\sigma Y}[c,j,k]$$

$$\left(\hat{r}_Y[c,j,k] \cdot T_2\left[modelID, id[c], \sigma_Y\left[c, \left\lfloor\frac{i}{8}\right\rfloor, \left\lfloor\frac{j}{8}\right\rfloor\right]\right]\right)/2^{16} \quad (9)$$
$$\rightarrow \hat{r}_Y[c,j,k]$$

where $id[c] = GRFS_Y[c] + 2 \cdot rvs\_enable\_flag[0]$ and $0 \leq c,i,j < 160, \left\lceil\frac{H}{16}\right\rceil, \left\lceil\frac{W}{16}\right\rceil$. The $GRFS_Y[c]$ are flags with values 0 or 1 that are signalled in the picture header when grfs_enable_flag[0] is true.

The entries of tables $T_1[4,4,3968]$ and $T_2[4,4,3968]$ are specified in JPEG AI Part 1, which additionally specifies how the entries can be calculated on the fly to provide flexibility in implementations. In the processing of secondary component with RVS, the same tables $T_1$ and $T_2$ are used.

### H. Latent Scaling Before Synthesis (LSBS)

LSBS is a coding tool similar to RVS. The inputs of the LSBS process are the residual tensors $\hat{r}_Y$ and $\hat{r}_{UV}$ after entropy decoding and the latent tensors $\hat{y}_Y$ and $\hat{y}_{UV}$ after the latent sample reconstruction process. Additionally, the average variance tensor $\sigma_Y$ and $\sigma_{UV}$ used in RVS are utilized in LSBS. In this subsection, only the application of LSBS to the primary component is described, as the application to the secondary component is identical.

*lsbs_enable_flag[comp]* is signalled in the picture header to control application of LSBS on the primary and secondary components (comp=0 and comp=1 indicating the primary and secondary components, respectively). The LSBS process on the primary component starts by calculating the tensor $\mu_Y[c,i,j] = \hat{y}_Y[c,i,j] - \hat{r}_Y[c,i,j]$, then $\hat{y}_Y$ is updated according to (10).

$$\hat{y}_Y[c,i,j] + \left(\hat{r}_Y[c,i,j] \cdot T_R\left[modelID, \sigma_Y\left[c, \left\lfloor\frac{i}{8}\right\rfloor, \left\lfloor\frac{j}{8}\right\rfloor\right]\right] \right. \\ + \mu_Y[c,j,k] \\ \left. \cdot T_P\left[modelID, \sigma_Y\left[c, \left\lfloor\frac{i}{8}\right\rfloor, \left\lfloor\frac{j}{8}\right\rfloor\right]\right] + 2^{12}\right) \\ \gg 13 \to \hat{y}_Y[c,i,j] \quad (10)$$

with $0 \leq c,i,j < 160, \left\lceil\frac{H}{16}\right\rceil, \left\lceil\frac{W}{16}\right\rceil$.

The entries of tables $T_P[4,3968]$ and $T_R[4,3968]$ are specified in JPEG AI Part 1, which additionally specifies how the entries can be calculated on the fly to provide flexibility in implementation. In the LSBS processing of the secondary component, the same tables $T_P$ and $T_R$ are used.

### I. Rate adaptation and quality map

JPEG AI provides three methods to adjust the quality of the reconstruction. The first is the most straightforward and is simply the selection of one model (i.e. the weights and biases) among four models trained for specific Lagrange multipliers $\beta_{train}$. To have finer rate adaptation a channel-wise map based on the concept of gain units is also available along with a spatial quality map to allow flexible spatial bit allocation.

A set of four models are specified in JPEG AI Part 1, which are stored in ONNX format [41] and can be downloaded from the link in [18]. The four model parameters are designed for operation at four different rate points, resulting in a wide range of adaptation possibilities. The *modelID* syntax element is signaled in the picture header which specifies the index of the model parameters that need to be used in decoding a codestream. Roughly speaking, a Bits Per Pixel (BPP) difference of 20× is possible between the lowest quality and the highest quality model parameters, though the exact difference in BPP depends on the content and the encoder implementation.

The *gain unit* and *3D gain unit* processes of rate control involve adjustment of the quantized residual ($\hat{r}_Y$ and $\hat{r}_{UV}$) and the variance tensors ($I_{\sigma Y}$ and $I_{\sigma UV}$) at the decoder, in a way similar to quantization control. The adjustment of the tensors follows the following equations.

$$m_{\log}[comp,c,i,j] \\ = betaDisplacementLog[comp] \quad (11) \\ + m_{ref}[modelID, comp, c]$$

$$m_{log}[comp,c,i,j] + Gain3d[i,j] \to m_{log}[comp,c,i,j] \quad (12)$$

$$m^{-1}[comp,c,i,j] = e^{-\frac{(m_{\log}[comp,c,i,j] \cdot step)}{2^{sigmaPrecision}}} \quad (13)$$

$$\hat{r}_Y[c,i,j] \cdot m^{-1}[0,c,i,j] \to \hat{r}_Y[c,i,j] \\ \hat{r}_{UV}[c,i,j] \cdot m^{-1}[1,c,i,j] \to \hat{r}_{UV}[c,i,j] \quad (14)$$

$$I_{\sigma Y}[c,i,j] + m_{\log}[0,c,i,j] \to I_{\sigma Y}[c,i,j] \\ I_{\sigma UV}[c,i,j] + m_{\log}[1,c,i,j] \to I_{\sigma UV}[c,i,j] \quad (15)$$

where, $0 \leq i,j < \left\lceil\frac{H}{16}\right\rceil, \left\lceil\frac{W}{16}\right\rceil$ and $0 \leq c < 160$ for primary component and $0 \leq c < 96$ for secondary component.

An adjustment tensor $m_{\log}$ is constructed firstly as in (11), using $betaDisplacementLog[comp]$ (comp=0 for primary and 1 for secondary component). $betaDisplacementLog$ is the parameter controlling the rate, signaled in the picture header. The quality of the primary and secondary components of the picture can be adjusted independently by signaling different values for $betaDisplacementLog[comp]$.

If *gain_3D_enable_flag* in the picture header is true, $m_{\log}$ is modified in (12) according to the $Gain3d[i,j]$ tensor, whose sample values are derived from the quality map codestream segment. The $Gain3d$ tensor allows adaptation of the quantization parameters spatially, facilitating applications such RoI coding.

After the construction of the $m_{\log}$ tensor, quantized residual and variance tensors are modified according to (13), (14) and (15). The parameters *step* and *sigmaPrecision* and the tensor $m_{ref}$ are predefined, comprising 12-bit signed integer values. The exponentiation operation in (13) can be replaced with a lookup-table in practical implementations. The equations (11)-(15) describe the process that needs to be performed at the decoder. At the encoder, additionally to the above, the unquantized residual tensors $r_Y$ and $r_{UV}$ should also be scaled in an opposite way to the scaling process in (14).[42] provides a deeper analysis of rate adaptation in JPEG AI together with an example encoder algorithm.

### J. Tiling and spatial random access

JPEG AI provides two schemes for partitioning an image, targeting different processes in the decoding pipeline. The first mechanism, named synthesis transform tiling, allows tiling of reconstructed latent tensors ($\hat{y}_Y$ and $\hat{y}_{UV}$) for processing with the synthesis transform. It allows obtaining the reconstruction in a patch-by-patch manner and eliminating the need to process the whole tensor at once. The primary goal of the synthesis transform tiling is to reduce and limit the peak memory usage when processing large images, which may be necessary in some decoder implementations.

The second partitioning scheme, named region partitioning, targets the entropy decoding and latent reconstruction processes. Region partitioning allows dividing and encapsulating the residual stream ($r_Y$ − stream and $r_{UV}$ − stream) into a separate codestream segments, wherein each residual segment can be decoded to obtain the corresponding region in the reconstructed latent tensors ($\hat{y}_Y$ and $\hat{y}_{UV}$). The partitioning of the residual stream paves the way for various use cases where a partial decoding of a whole picture is performed, including 360-degree imaging, virtual reality applications, RoI coding to name a few.

Both schemes can be used together to meet application requirements and to manage peak memory usage. In the picture header two enable flags, *synthesis_tile_enable[comp]* and *region_partitioning_flag*, are signaled to indicate usage of synthesis transform tiling and region partitioning, respectively. In addition, the flag *region_residual_in_its_own_substream_flag* determines whether regions are coded independently and that the residual stream is encapsulated in separate codestream segments. When *region_residual_in_its_own_substream_flag* is equal to true, it is required that each region shall contain an integer multiple number of synthesis transform tiles. Furthermore, the latent sample reconstruction (Subsection VI.D) and synthesis transformation (Subsection VI.F) processes are restricted from using samples from neighboring regions. An example partitioning of a picture into regions and synthesis transform tiles in this case is provided in the Appendix.

*K. Processing of arbitrary image sizes*

Two methods are used to handle images with arbitrary sizes, namely the layer-based cropping mechanism and the display window mechanism. In the encoder, the input image signal goes through series of downsampling layers, in each one the vertical and horizontal size of the input tensor is reduced by half. When the sizes of the input tensor are not multiples of 2 before downsampling, padding is applied to extend the tensor in either or both directions by one sample. In the analysis transform and hyper encoder networks, there is a padding layer in front of each downsampling layer, 4 down sampling layer in analysis and 2 in hyper encoder network. This right-on-point padding approach reduces the redundant processing, as an intermediate tensor is extended at most by a single line of samples in horizontal and/or vertical direction.

The padding operation is reverted in the decoder, by the cropping layer after each upsampling layer in the synthesis, hyper decoder and hyper scale decoder networks. Each padding layer in the encoder networks has a corresponding cropping layer in the decoder networks to revert its impact.

Though the layer-based cropping mechanism allows handling images with arbitrary size while reducing redundant processing, it was noticed during the standardization process that it cannot be gracefully supported by some device implementations, since it requires changing the intermediate tensor size between two layers of a subnetwork. To address this issue, the display window mechanism is included in JPEG AI. Accordingly, the image to be processed might be larger than the image to be displayed, allowing the padding of the original image before compression and cropping of the reconstructed image after decompression. This mechanism provides an option to pad an original image to be a multiple of 64 samples in both the vertical and horizontal dimensions. In this case, cropping operation of the intermediate tensors are not invoked, potentially increasing the decoding speed in some implementations. Two parameters *diff_display_img_width* and *diff_display_img_height* control the display window size and are signaled in the picture header.

*L. Progressive Decoding*

Progressive decoding allows partial decoding of the codestream, and construction of a lower resolution or a preview image in ways that are not normatively specified in the standard specification. It is particularly useful in scenarios including low transmission bandwidth or low device capabilities. In JPEG AI, progressive decoding can be achieved simply by setting the part of the residual tensors $\hat{r}_Y$ and $\hat{r}_{UV}$ to be zero and continuing the latent sample reconstruction and synthesis transform processes with partially filled residual tensors.

*M. Post processing filters*

Four post processing filters are specified in JPEG AI Part 1. The main motivation of these filters is to enhance the quality of the decoded image after reconstruction, especially the color, detail and contrast information that can be lost due to the quantization and the reconstruction processes. Two of the four filters, EFE (Enhancement Filtering Extensions) linear and EFE nonlinear filters modify only the secondary component, the Inter-Channel Correlation Information (ICCI) filter modifies both the primary and secondary components, and the LEF (Luma Edge Filter) modifies only the primary component. The post processing filters are optional, and the decoder can skip their application even if they are enabled by the syntax. The control parameters of post processing filters are located in the tools header.

VII. EXPERIMENTAL RESULTS

*A. Experimental settings*

To evaluate the performance of JPEG AI, both objective and subjective tests were conducted on several datasets, including the JPEG AI test dataset [23], the JPEG AI synthetic dataset [23], the Kodak dataset [43], and the CLIC 2024 validation dataset [44]. The JPEG AI test dataset comprises 50 natural images with resolutions ranging from 1K to 4K and serves as the common test condition in standard development. Additionally, we evaluated performance on the JPEG AI synthetic dataset, which includes 36 images featuring animation, screen, and game content. Consistent with prior research, we also performed experiments on the Kodak dataset and the CLIC 2024 validation dataset.

In the experiments, the VVC Intra was used as the anchor. FFmpeg [46] is employed to convert the dataset from PNG format to YUV format. Subsequently, VTM11.1 [47] is used to generate bitstreams and reconstructed pictures for VVC [48] which is one of the anchors specified in JPEG AI CTTC. For JPEG AI, VM7.0 [49] is used in the tests and different encoder and decoder configurations were evaluated.

Following the result reporting procedure established in the JPEG AI CTTC, the Bjøntegaard Delta rate (BD-Rate) [45] was calculated to demonstrate the bitrate saving of JPEG AI using the seven quality metrics introduced in Section II. For measuring complexity, kilo Multiply-Accumulate Operations per pixel (kMAC/pxl) is used to quantify the number of operations performed in a given test configuration as specified in [23]. Since the JPEG AI common test conditions used in the experiments contain images with varying resolutions, decoding time was normalized using milliseconds per megapixel, representing the average time required to decode a 1,000 × 1,000 pixel image patch. In the tests, decoding is performed on NVIDIA Tesla V100 Volta GPU Accelerator with 32GB of memory and Intel Xeon Platinum 8336C CPU @ 2.30GHz. The commands in running JPEG VM and VTM software are provided in subsections I and J of the Appendix.

*B. Results on the CTTC dataset*

First, the performance is evaluated, including compression efficiency and computational complexity, on the JPEG AI test dataset. Results in Table IV analyze the performance of several combinations of analysis and synthesis transforms. In [18] two analysis transforms, and three synthesis transforms are specified, which are the main components differentiating the different profiles of JPEG AI. In this test, two sets of bitstreams are generated using two analysis transforms (denoted encoderID equal to 0 and 1), each of which are decoded using three synthesis transforms (denoted decoderID equal to 0, 1 and 2). To demonstrate the reconstruction quality of JPEG AI, comparative results are provided in Table III. The verification model of JPEG AI [49] is compared with reference software of VVC [47] Intra as benchmark. When encoderID is equal to 0, the JPEG AI outperforms VVC Intra in five of the seven quality metrics (MS-SSIM, FSIM, NLPD, IW-SSIM and VMAF) in terms of BD-rate by a large margin. BD-rate gains of the seven quality metrics are averaged and the result is presented in the "AVG" column. According to the average BD-rate results, the three different decoders (decoderID=0, 1 and 2) outperform VVC Intra by 16.0%, 20.2% and 21.1%. When the bitstream is generated with the second encoder, the three decoders outperform VVC Intra by 13.9%, 19.7%, and 27% in terms of average BD-rate. The encoder that is used in the bitstream generation depends on the target decoders. The two encoders specified in the JPEG AI reference software and JPEG AI core coding systems are not a normative part of the standard. The implementors are free to develop their own encoders suited to their application needs.

TABLE IV - TOOLS DISABLE TEST (ANCHOR IS VTM)

| TEST | AVG | MS-SSIM | VIF | FSIM | NLPD | IW-SSIM | VMAF | PSNR-HVS | kMAC/pxl | Decoding time (ms/megapixel) |
|---|---|---|---|---|---|---|---|---|---|---|
| All on | -20.2% | -33.0% | 1.4% | -26.9% | -17.3% | -29.1% | -34.9% | -1.9% | 27.7 | 266 |
| RVS off | -18.0% | -33.1% | 0.8% | -20.2% | -14.8% | -29.0% | -28.8% | -0.7% | 27.7 | 249 |
| LSBS off | -19.8% | -33.1% | 1.8% | -27.4% | -17.5% | -29.1% | -30.5% | -2.7% | 27.6 | 249 |
| LEF off | -19.9% | -33.2% | 1.2% | -27.3% | -18.1% | -29.0% | -29.0% | -3.7% | 27.7 | 251 |
| ICCI off | -20.0% | -32.8% | 1.7% | -26.9% | -17.2% | -29.0% | -34.1% | -2.0% | 23.1 | 245 |
| EFE nonlinear off | -20.4% | -33.3% | 1.2% | -26.2% | -17.6% | -29.4% | -35.1% | -2.1% | 27.7 | 246 |
| EFE linear off | -20.4% | -33.3% | 1.1% | -26.1% | -17.6% | -29.4% | -35.2% | -2.2% | 28.6 | 250 |

TABLE V PERFORMANCE EVALUATION ON KODAK DATASET (ANCHOR IS VTM)

| TEST | AVG | MS-SSIM | VIF | FSIM | NLPD | IW-SSIM | VMAF | PSNR-HVS |
|---|---|---|---|---|---|---|---|---|
| JPEG AI decoderID=0 | -7.5% | -29.8% | 18.1%% | -19.7% | -0.2% | -24.1% | -22.3% | 25.3% |
| JPEG AI decoderID=1 | -12.9% | -32.1% | 11.4% | -22.9% | -6.3% | -26.8% | -28.4% | 14.5% |
| JPEG AI decoderID=2 | -21.1% | -37.3% | 0.0% | -28.8% | -15.6% | -32.0% | -38.3% | 4.4% |

TABLE VI PERFORMANCE EVALUATION ON CLIC 2024 VALIDATION DATASET (ANCHOR IS VTM)

| TEST | AVG | MS-SSIM | VIF | FSIM | NLPD | IW-SSIM | VMAF | PSNR-HVS |
|---|---|---|---|---|---|---|---|---|
| JPEG AI decoderID=0 | -12.1% | -25.7% | 22.6% | -30.8% | -7.6% | -25.0% | -25.4% | 7.3% |
| JPEG AI decoderID=1 | -16.8% | -28.4% | 15.4% | -34.6% | -12.2% | -27.9% | -32.0% | 1.9% |
| JPEG AI decoderID=2 | -24.9% | -34.5% | 2.8% | -42.3% | -19.9% | -33.5% | -40.7% | -6.3% |

Table IV two columns, kMAC/pxl and decoding time, are presented regarding the complexity. In terms of kMAC/pxl, the decoders with IDs 0, 1 and 2 have 13, 28 and 215 kMAC/pxl complexity. The decoding time of $1k \times 1k$ pixel patch is between 246 and 332 ms/megapixel for different encoder-decoder combinations.

When assessing complexity, the capabilities of the target device are the ultimate determining factor. In the case of a high-performance device such as the Tesla V100 GPU that is used in the tests, the decoding time does not increase much even if the kMAC/pxl is increased significantly among the different decoders. In the case of a handheld device, however, different considerations such as power consumption and memory usage play a more significant role in the design considerations. To facilitate deployment in a wide range of devices, the JPEG AI employs profiles where reconstructions with multiple decoders are deemed compliant to the standard, which provides the implementers with a spectrum of complexity vs. coding gain tradeoff.

To verify its generalization performance, the JPEG AI codec is evaluated on the Kodak dataset and the CLIC 2024 validation dataset. The results are presented in Table V and Table VI Compared with the results on the JPEG AI dataset, the BD-rate shows a negligible amount of variation. On these datasets JPEG AI provides roughly between 7% and 25% coding gain depending on the selected decoder.

*C. Ablation study*

We also conduct a tool-off test to verify the effectiveness of the switchable tools as part of the ablation study, as shown in Table IV. In the experiment the encoder with encoderID=0 was used to generate the bitstreams, the decoder with decoderID=1 for decoding, and all tools were enabled to obtain the result labeled 'All on.' Then, each one of the tools was disabled individually to observe the performance difference, which allows to assess the impact of each JPEG AI tool. From the table, it is possible to observe that the RVS tool provides 2.2% average coding gain. The gains provided by LSBS and LEF are similar, with each contributing around 0.4% and 0.3%. Additionally, ICCI provides an average gain of 0.2%. Unlike the above tools, EFE nonlinear and EFE linear lead to a slight loss due to the additional side information in the bitstream. However, the benefit of the EFE linear and EFE non-linear filter is the improvement obtained in the chroma PSNR, around 12% and 8% respectively. Though chroma PSNR is not among the metrics that are used in the development of JPEG AI, the significant amount of coding gain for the color component justified their inclusion in Part 1 of JPEG AI.

### D. Subjective examples

To demonstrate the reconstruction quality of JPEG AI, comparative results are provided in Fig. 3. Reference software of JPEG AI [49] is compared with reference software of VVC [47] Intra as benchmark. Both natural content and synthetic content images were selected from the JPEG AI CTTC dataset.

Each image is coded at two different rate points, approximately 0.08 BPP and 0.3 BPP. Low bitrates are selected to demonstrate the difference in visual quality, since the coding artefacts become less noticeable with increasing bitrate. In the case of JPEG AI, the same codestream is decoded with two different JPEG AI decoders: the BOP (Base Operating Point) decoder and HOP (High Operating Point) decoder. The post processing filters that are described in Subsection VI.M are disabled. At higher bitrates, the differences between reconstructed images become increasingly difficult to discern, as coding artifacts diminish with the increase in bitrate.

For natural content images, it is observed that JPEG AI can preserve more textual details compared to VVC Intra, especially at the lowest rate point. When the codestream is decoded with BOP or HOP decoder, the reconstructions exhibit superior visual quality compared to VVC Intra. Though the difference between the BOP and HOP reconstructions is less apparent, it can be observed that the colors are more vibrant and closer to the original in the case of the latter.

An example of synthetic content image is also provided at similar bitrates. In this case, it is possible to observe that HOP decoder can produce higher fidelity compared to BOP decoder, though they both struggle with sharp straight lines, although overall visual quality is on par with VVC Intra. One limitation of JPEG AI is that its performance is not consistent in synthetic content, it can be observed that especially in the case of screen captured images comprising letters, the reconstruction quality is lower than VVC in the areas containing letters. The synthetic content coding has not been the main focus in the development of first version of JPEG AI due to time limitations. Improving the screen content coding will likely be one of the focus points in the future versions of JPEG AI.

## VIII. PROFILING

Part 2 of JPEG AI specifies profiles and levels. JPEG AI utilizes a nested profile structure comprising stream profiles and their associated decoder profiles. A stream profile defines a subset of codestream syntax and their admissible values. Each stream profile might be associated with one or more decoder profiles, where each decoder profile specifies a subset of decoder tool set. A codestream conforming to a specific stream profile is decodable by all associated decoder profiles, enabling reconstructions optimized for different purposes or levels of decoding complexity. The stream and decoder profile identifiers are included in the picture header.

In draft version of Part 2 [19], one stream profile, the Main stream profile, and three decoder profiles, Main@Simple, Main@Base, Main@High, are defined. These decoder profiles are differentiated by the supported synthesis transforms. In the Main@Simple profile, only the least complex synthesis transform, identified by decoderID=0, is supported. The Main@Base profile allows both decoderID 0 and 1, while the Main@High profile supports all three synthesis transforms defined in Part 1. In the Main stream profile, the application of post processing filters detailed in Subsection VI.M is not mandatory.

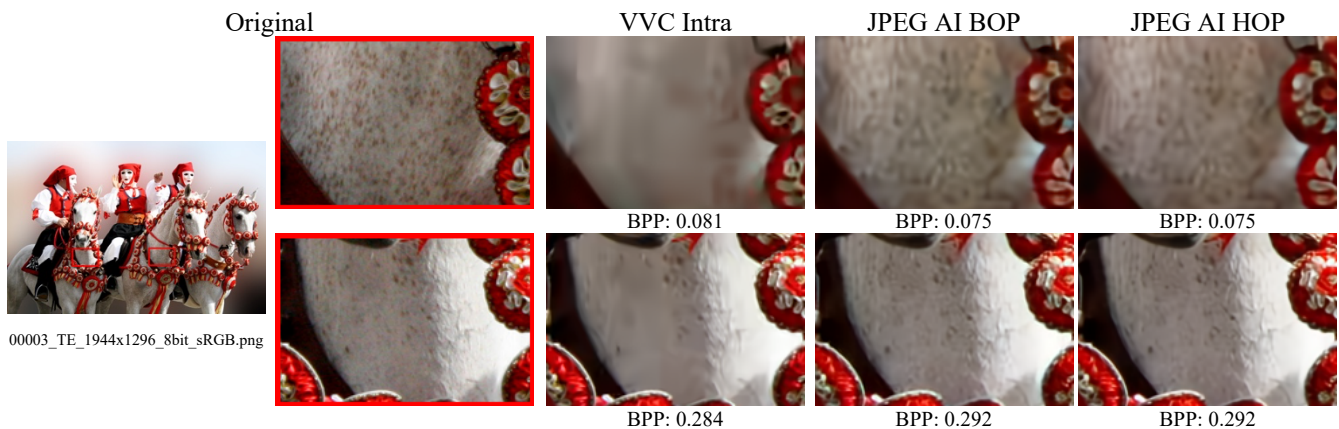

00003_TE_1944x1296_8bit_sRGB.png

| Original | VVC Intra | JPEG AI BOP | JPEG AI HOP |
| --- | --- | --- | --- |
| | BPP: 0.081 | BPP: 0.075 | BPP: 0.075 |
| | BPP: 0.284 | BPP: 0.292 | BPP: 0.292 |

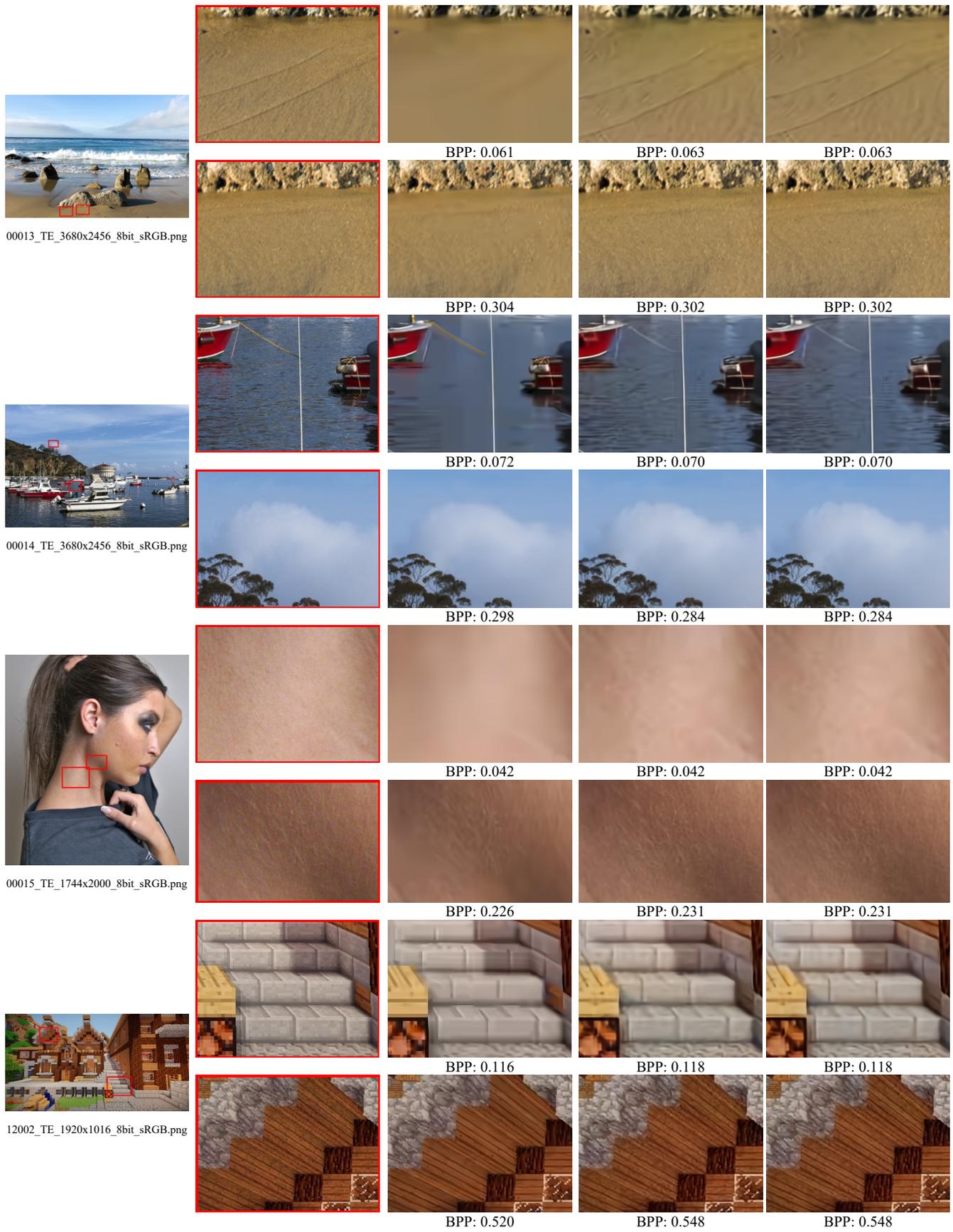

Figure 3. Subjective comparison results using JPEG AI BOP, JPEG AI HOP compared to VVC.

## IX. Conformance specification

Part 4 of JPEG AI specifies the requirements for generating a JPEG AI-compliant reconstruction. Specifically, it defines a testing suite for each profile and level pair specified in Part 2. If an implementation successfully passes all the designated tests, it is considered compliant with the corresponding profile and level. At the time of writing this paper, Part 4 of JPEG AI was still in a draft form, with the target finalization date set for October 2025.

## X. Reference software

JPEG AI provides reference software, known as the Verification Model (VM), to demonstrate the application of algorithms specified in the standard. The JPEG AI VM includes implementations for both the encoder and decoder used in image compression and decompression to facilitate research and development of standardization. JPEG AI VM also includes scripts for metric calculation and testing, enabling users to evaluate coding performance under CTTC. To further analyze the performance of each component in JPEG AI, profiling tools are provided to assess processing time and the impact of specific tools. JPEG AI Part 3 [21] provides the reference software and instructions on its usage.

## XI. JPEG AI File format

For enabling the use of JPEG AI coded images and image sequences in internet-based and other applications, JPEG AI Part 5 was developed, specifying the container file formats for JPEG AI codestreams based on the ISO Base Media File Format (ISOBMFF) [50] and the High Efficiency Image File format (HEIF) [51]. Specifically, Annex A of JPEG AI Part 5 specifies the use of JPEG AI coding for timed sequences of images within files based on ISOBMFF, denoted Motion JPEG AI. The Motion JPEG AI file format is designed to contain one or more motion sequences of JPEG AI compressed images, with their timing. Motion JPEG AI is a flexible format, permitting a wide variety of usages, such as editing, display, interchange, and streaming. JPEG AI Part 5 Annex B specifies a format to encapsulate JPEG AI images, image collections, and image sequences in HEIF, including file brands for a single image and an image collection as well as image sequences.

## XII. Conclusion

The JPEG AI standard represents a significant milestone as the first international standard aimed at harnessing advancements in learning-based image compression. By incorporating a variety of design features, it enables practical implementations across a wide range of devices and platforms on the market. Additionally, the standard integrates a diverse set of tools to support a broad spectrum of use cases and applications, including thumbnail generation, transcoding, virtual reality, 360-degree imaging, region-of-interest coding, wide dynamic range imaging, and decoder complexity scalability, among others. This paper has provided an overview of these capabilities and design elements, highlighting the standard's potential to address the evolving needs of modern image compression. For further details, the readers are referred to the technical specifications of JPEG AI in [18][19][20][21][22].

## XIII. Future Directions

The first version of JPEG AI was developed with simplicity and implementability as primary goals, favoring cautious design choices to ensure broad applicability. As the first learning-based image coding standard, it provides foundational and limited support for certain use cases and requirements. For instance, JPEG AI offers only basic support for synthetic content coding, as it does not include dedicated coding tools or model parameters specifically optimized for this purpose. While certain design adaptations have been introduced to maintain acceptable reconstruction quality for synthetic content, its compression performance in such cases generally falls short compared to traditional image coding standards.

Additionally, features such as bit-exact picture reconstruction and lossless coding were deliberately excluded from the first version to prioritize greater implementation flexibility. Such requirements, along with improved support for different use cases and further improvements in compression efficiency, are anticipated to be addressed in a potential future version of JPEG AI. Moreover, in recent years, many research directions have been exploited to further enhance the compression efficiency of learning-based image compression, such as solution with implicit neural networks (online training), diffusion models and transformers-based architecture. Therefore, it is necessary to continue to monitor these developments in the future to understand if there is clear evidence of advances in terms of compression efficiency and complexity. As the devices become more powerful it is expected that such type of solutions could play a significant role in new version of the JPEG AI standard.

Another promising direction for future research is the consumption of images by machines. Today, visual content is not only consumed by humans but also increasingly by machines that perform image processing and computer vision tasks. The JPEG AI latent representation, generated by the analysis transform and available in its quantized form at the decoder, can be directly leveraged for a wide range of visual analysis tasks, such as object detection, recognition, and semantic segmentation, as well as enhancement tasks like super-resolution, denoising, and color correction. Using the latent representation instead of the decoded (and thus lossy) image allows these tasks to be performed with lower computational complexity, and in some cases, with higher accuracy, particularly at lower quality settings. While initial research in this area has already shown promising results, the next version of JPEG AI aims to further address these use cases, fostering a more efficient and sustainable ecosystem around its visual representation.

## Acknowledgment

The authors would like to thank the experts of ITU-T, ISO/IEC JPEG and MPEG for their contributions in the development of the JPEG AI standard.

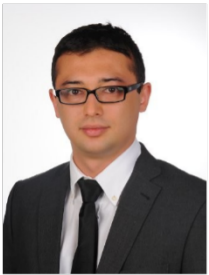

**Semih Esenlik** received his B.S. degree on Electrical and Electronics Engineering from Bogazici University of Istanbul in 2008 and his M.S. degree from Technical University of Munich on Telecommunications Engineering in 2010. From 2010 to 2013 he worked as Research Engineer in Panasonic R&D Center of Germany, in the Multimedia Coding and Standardization Group. Between 2013 and 2016, he worked as technical product manager at Turkcell in Turkey. In 2016 he started working in Huawei R&D Center of Germany as principal research engineer and shortly after became team leader of video coding standardization team. He has actively participated in the development of HEVC, EVC and VVC codecs, is among the main contributors of decoder side motion vector refinement and geometric partitioning mode technologies in VVC. He was the primary chair of core experiment on decoder side motion vector derivation and refinement during the development of VVC. In 2021 he joined Bytedance at United States as principal researcher and started working on development of neural networks-based image codec and contributed to the CfP issued by JPEG AI. The response submitted by Bytedance is selected one of the two candidates to be integrated into "Verification Model under Consideration" of JPEG AI. He is currently an editor of JPEG AI.

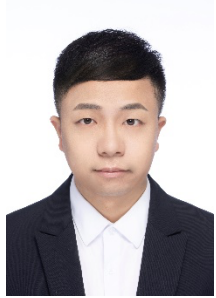

**Yaojun Wu** received the B.S. degree from the School of Electronics and Information Engineering, Anhui University, Hefei, China, in 2018, and the M.S. degree from the School of Information Science and Technology, University of Science and Technology of China, Hefei, in 2021. He is currently working in the standardization team of Bytedance Inc. as a senior engineer. He has been an active contributor in JPEG AI and IEEE 1857.11. During these standards, he serves as the software coordinator for JPEG AI. His research interests include learning-based image and video compression, representation learning, and quality assessment.

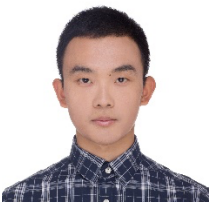

**Zhaobin Zhang** received the B.S. and M.S. degrees in mechanical electronic engineering from the Huazhong University of Science and Technology, Wuhan, China in 2012 and 2015, respectively. He obtained the Ph.D. degree in Computer Science and Electrical Engineering, University of Missouri-Kansas City. He is currently a Research Scientist with the Multimedia Lab, Bytedance Inc., San Diego, CA, USA. His research interests include learning-based image and video compression. He was an active contributor to JPEG AI and IEEE 1857.11 on learning-based image compression standards.

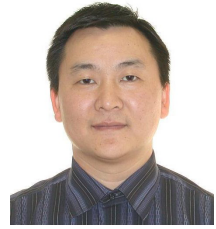

**Ye-Kui Wang** received his BS degree in industrial automation in 1995 from Beijing Institute of Technology, and his PhD degree in information and telecommunication engineering in 2001 from the Graduate School in Beijing, University of Science and Technology of China.

He is currently a Principal Scientist at Bytedance Inc., San Diego, CA, USA. His earlier working experiences and titles include Chief Scientist of Media Coding and Systems at Huawei Technologies, Director of Technical Standards at Qualcomm, Principal Member of Research Staff at Nokia Corporation, etc. His research interests include video coding, storage, transport and multimedia systems.

Dr. Wang has been an active contributor to various multimedia standards, including video codecs, media file formats, RTP payload formats, and multimedia streaming and application systems, developed by various standardization organizations including ITU-T VCEG, ISO/IEC MPEG, JVT, JCT-VC, JCT-3V, JVET, 3GPP SA4, IETF, AVS, DVB, ATSC, and DECE. He has been chairing the development of OMAF at MPEG, and has been an editor for numerous standards, including VVC, VSEI, OMAF, HEVC, MVC, later versions of AVC, a recent version of ISO base media file format, JPEG AI file format, VVC file format, HEVC file format, layered HEVC file format, SVC file format, a recent version of CMAF, a recent amendment of DASH, ITU-T H.271, RFC 6184, RFC 6190, RFC 7798, RFC 9328, 3GPP TR 26.906, and 3GPP TR 26.948.

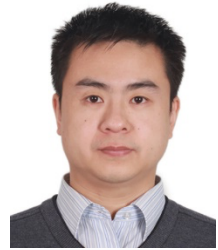

**Kai Zhang** (M'17, SM' 20) received the B.S. degree in computer science from Nankai University, Tianjin, China, in 2004. In 2011, he received the Ph.D. degree in computer science from the Institute of Computing Technology, Chinese Academy of Sciences, Beijing, China. From 2011 to 2012, he worked as a researcher in Tencent Inc. Beijing, China. From 2012 to 2016, he worked as a team manager in Mediatek Inc. Beijing, China, leading a research team to propose novel technologies to emerging video coding standards. From 2016 to 2018, he worked in Qualcomm Inc. San Diego, CA, still focusing on video coding standardization. Now, he is leading the standardization team in Bytedance Inc. San Diego, CA. Dr. Zhang' research interests include video/image compression, coding, processing and communication, especially video coding standardization. From 2006, he has contributed more than 500 proposals to JVT, VCEG, JCT-VC, JCT-3V, JVET and AVS, covering many important aspects of major standards such as H.264/AVC, HEVC, 3D-HEVC, VVC and AVS-1,2,3. He has 800+ granted or pending U.S. patents applications. Most of these patents are essential to popular video coding standards. During the development of VVC, Dr. Zhang co-chaired several core experiments and branch of groups. Currently, Dr. Zhang serves as a coordinator of the reference software known as ECM in JVET, to explore video coding technologies beyond VVC. Dr. Zhang has co-authored 100+ papers and reviewed 80+ papers on top-tier journals/conferences. He was a TPC member for VCIP 2018 and DCC 2024. He was an organizer of the Grand Challenge on Neural Network-based Video Coding in ISCAS 2022/2023/2024. Now he is the AE of IEEE T-CSVT and IET-IP.

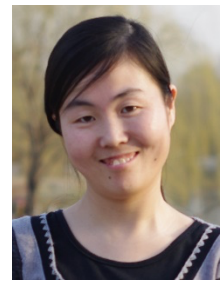

**Li Zhang** (M'07-SM'20) received the B.S. degree in computer science from Dalian Maritime University, Dalian, China, in 2003, and the Ph.D. degree in computer science from the Institute of Computing Technology, Chinese Academy of Sciences, Beijing, China, in 2009.

From 2009 to 2011, she held a post-doctoral position at the Institute of Digital Media, Peking University, Beijing. From 2011 to 2018, she was a Senior Staff Engineer, with the Multimedia R&D and Standards Group, Qualcomm, Inc., San Diego, CA, USA. She is currently the Lead of the Multimedia Lab, Bytedance Inc., San Diego, CA, USA. Her research interests include 2D/3D image/video coding, video processing, and transmission.

She was a Software Coordinator for Audio and Video Coding Standard (AVS) and the 3D extensions of High Efficiency Video Coding (HEVC). She has authored 600+ adopted standardization contributions, 800+ granted US patents, 150+ technical articles in related book chapters, journals, and proceedings in image/video coding and video processing. She has been an active contributor to H.266/Versatile Video Coding, AVS, IEEE 1857, extensions of H.264/AVC and H.265/HEVC, MPEG G-PCC (Geometry based Point Cloud Compression) and JPEG AI. During the development of those video coding standards, she co-chaired several ad hoc groups and core experiments. She has been appointed as an Editor of AVS, the Main Editor of the Software Test Model for 3DV Standards. She organized/co-chaired multiple special sessions and grand challenges at various conferences/journals. She is a Senior Member of IEEE, serves as associate editor in IEEE Transactions on Circuits and Systems for Video Technology (T-CSVT), Area Chair of ICME 2025, Publicity Subcommittee Chair of the Technical Committee member of Visual Signal Processing and Communications in IEEE CAS Society (VSPC TC)(2021-2024), Multimedia Signal Processing Technical Committee (MMSP TC) (2025-2028), reviewer of various top conferences including CVPR'25 IJCAI'25 ICML'25, etc. al.

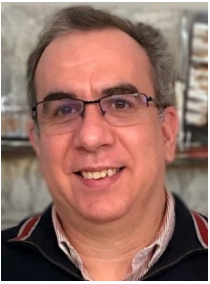

**Joao Ascenso** (Senior Member, IEEE) received the E.E., M.Sc., and Ph.D. degrees in electrical and computer engineering from the Instituto Superior Técnico (IST), Universidade Técnica de Lisboa, Lisbon, Portugal, in 1999, 2003, and 2010, respectively. He is currently an Associate Professor with the Department of Electrical and Computer Engineering, IST, and a member of the Instituto de Telecomunicações. He has published more than 150 papers in international conferences and journals. His current research interests include visual coding, quality assessment, coding and processing of 3D visual representations, coding for machines, super-resolution, denoising among others. He was an associate editor of IEEE Transactions on Image Processing, IEEE Signal Processing Letters and IEEE Transactions on Multimedia and guest editor of IEEE Access and IEEE Transactions on Circuits and Systems for Video Technology. He received three Best Paper Awards at PCS 2015, ICME 2020, and MMSP 2024 and has served as Technical Program Chair and in other organizing committees roles of major international conferences, including IEEE ICIP, PCS, EUVIP, ICME, MMSP and ISM.

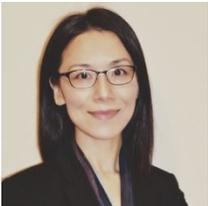

**Shan Liu** (Fellow, IEEE) received the B.Eng. degree in electronic engineering from Tsinghua University, the M.S. and Ph.D. degrees in electrical engineering from the University of Southern California, respectively. She is a Distinguished Scientist and General Manager at Tencent. She was formerly Director of Media Technology Division at MediaTek USA. She was also formerly with MERL and Sony, etc. She has been a long-time contributor to international standardization with many technical proposals adopted into various standards such as VVC, HEVC, OMAF, DASH, MMT and PCC, and served as a Project Editor of ISO/IEC | ITU-T H.266/VVC standard. She is a recipient of ISO&IEC Excellence Award, Technology Lumiere Award, USC SIPI Distinguished Alumni Award, and two-time IEEE TCSVT Best AE Award. She currently serves as Associate Editor-in-Chief of IEEE Transactions on Circuits and Systems for Video Technology and Vice Chair of IEEE Data Compression Standards Committee. She also serves and has served on a few other Boards and Committees. She holds more than 600 granted US patents and has published more than 100 peer-reviewed papers and one book. Her interests include audio-visual, volumetric, immersive and emerging multimedia compression, intelligence, transport and systems.